\documentclass[aps,prl,twocolumn,groupedaddress,showpacs]{revtex4}

\usepackage{graphicx}
\usepackage[ansinew]{inputenc}
\usepackage{array}
\usepackage{color}
\usepackage{amsmath}
\usepackage{amsxtra}
\usepackage{amstext}
\usepackage{amssymb}
\usepackage{latexsym}
\usepackage{dsfont} 
\begin{document}


\title{Detecting level crossings without looking at the spectrum}


\author{M. Bhattacharya and C. Raman}
\affiliation{School of Physics, Georgia Institute of Technology,
Atlanta, Georgia 30332}


\date{\today}

\begin{abstract}

In many physical systems it is important to be aware of the
crossings and avoided crossings which occur when eigenvalues of a
physical observable are varied using an external parameter.  We have
discovered a powerful algebraic method of finding such crossings via
a mapping to the problem of locating the roots of a polynomial in
that parameter. We demonstrate our method on atoms and molecules in
a magnetic field, where it has implications in the search for
Feshbach resonances. In the atomic case our method allows us to
point out a new class of invariants of the Breit-Rabi Hamiltonian of
magnetic resonance. In the case of molecules, it enables us to find
curve crossings with practically no knowledge of the corresponding
Born-Oppenheimer potentials.

\end{abstract}


\maketitle

Curve crossing is a universal phenomenon with examples in many
branches of both pure and applied sciences. It is responsible for
electron transfer in proteins \cite{Marcus1985}, underlies stability
analysis in mechanical engineering \cite{Seyranianbook} and
determines Nash equilibria \cite{Petrosjanbook}. Mathematically, it
arises naturally in the subject of algebraic geometry
\cite{Basubook}. In physics, the discovery of a crossing between two
energy levels, for example, signals degeneracy in the energy
spectrum and the existence of an underlying symmetry in the problem.
Since the initial pioneering works by Hamilton
\cite{Hamconical1833}, von Neumann and Wigner
\cite{NWignersymm1929}, and Landau \cite{landau1932} and Zener
\cite{zener1932}, many interesting physical phenomena have been
associated with crossings. These include Berry's phase in adiabatic
quantum mechanics \cite{Berry1984}, the hidden symmetries of the
Hubbard model in condensed matter physics \cite{heilman1970}, and
the onset of quantum chaos in nonlinear dynamics \cite{Haakebook}.
In contemporary atomic and molecular physics an avoided crossing
that has come into intense focus is the Feshbach resonance. Such a
resonance provides unique experimental control over interactions in
a quantum degenerate atomic gas, realizing collapsing condensates
\cite{Cornish2000Wieman-85RbBEC}, ultracold molecules
\cite{greiner2003} and the crossover from BEC to BCS pairing in
degenerate Fermi gases \cite{zwierlein2003}.

In this Letter we present a versatile theoretical technique for
detecting the presence of level crossings in physical systems.  Our
technique is algebraic in nature, as the underlying Hamiltonian (or
other physical quantity) can often be represented as a
parameter-dependent matrix.  Our method is powerful and general, as
we extract essential information simply from the matrix elements
without having to compute the spectrum, and remarkably, works even
in cases where the Hamiltonian is not fully known. We demonstrate
our technique on atoms and molecules in a magnetic field. Apart from
being interesting in themselves, these physical systems are of
relevance to Feshbach resonances, a topic of current experimental
and theoretical concern \cite{Cornish2000Wieman-85RbBEC}. We present
specific results for the atom pair $^{23}$Na-$^{85}$Rb, which is of
current interest as a candidate for Feshbach resonances
\cite{pashov2005}, but our analysis can be readily extended to other
atoms.

The essence of our algebraic approach can be demonstrated using the
following simple example.  Let us consider a real symmetric
2$\times$2 matrix,
\begin{equation}
\label{eq:H2} M(T)= \begin{pmatrix}
        E_{1} &  V\\
        V & E_{2} \\
      \end{pmatrix}
\end{equation}
which may represent the Hamiltonian of a two-state system. The
notation implies that every matrix element is a function of the
tunable parameter $T$, which could be an external field.  The
eigenvalues $E$ of $M(T)$, labeled $\lambda_1$ and $\lambda_2$, may
be found from the characteristic polynomial $|M(T)-E| =
E^{2}+C_{1}E+C_{0}$, where $C_0 = E_1 E_2-V^2$ and $C_1 =
-(E_1+E_2)$.  However, since the eigenvalues are themselves roots of
this polynomial, i.e., $(E-\lambda_1)(E-\lambda_2) = 0$, the
coefficients may equivalently be written in terms of the
eigenvalues: $C_{0}= \lambda_{1} \lambda_{2}$ and
$C_{1}=-(\lambda_{1}+\lambda_{2})$. It should be noted that we did
not explicitly calculate the eigenvalues to arrive at this
conclusion.

Since we are interested in curve crossings, we would like to know if
the eigenvalues are degenerate, i.e., if $\lambda_1 = \lambda_2$, at
some value of the parameter $T$. To this end, we introduce the
\emph{discriminant} $\Delta \equiv (\lambda_{1}-\lambda_{2})^{2}$ of
the characteristic polynomial of $M(T)$ which can be written purely
in terms of its coefficients : $\Delta = C_{1}^{2}-4 C_{0}$.  If we
now choose a simple parametrization for $M(T)$, say $E_{1,2}=2T$ and
$V=T$, we find $\Delta = 4T^{2}$, a polynomial in $T$. The root of
$\Delta$, $T=0$ corresponds to the only crossing in the spectrum of
$M(T)$, as may be verified by explicitly calculating the eigenvalues
$\lambda_{1,2} = T,3T$. The point of this example is to show how use
of the discriminant maps the problem of finding spectral crossings
to one of locating polynomial roots and bypasses the need to find
eigenvalues.

An extension of this simple example to larger matrices and more than
one parameter leads to the study of multivariate polynomials, a
topic in algebraic geometry \cite{Basubook}. Assigning the task of a
rigorous exposition to a separate publication \cite{Crossingpra}, we
motivate the technique used in this Letter by analogy to the example
stated above. We consider a matrix $H(P)$, not trivially diagonal,
of dimension $n$ and depending polynomially on the set of parameters
$P = \{P_{1},..P_{N}\}$. The characteristic polynomial of $H(P)$ is
of degree $n$ in the eigenvalue $E$. The roots $\lambda_{1,2,..n}$
of this polynomial are used to define its discriminant
\begin{eqnarray}
\begin{array}{l}
\label{eq:Discdefn} D[H(P)]=
\displaystyle\prod_{i<j}^{n}(\lambda_{i}-\lambda_{j})^{2},\\
\end{array}
\label{eq:disc}
\end{eqnarray}
which is an invariant of $H(P)$. It follows from
Eq.(\ref{eq:Discdefn}) that $D[H(P)]$ does not change if the same
quantity is added to every eigenvalue $\lambda_{i}$; in fact it does
not even change if the same quantity is added to each diagonal term
of $H(P)$, since this corresponds to a shift of the entire spectrum.

Now as in the case of $\Delta$ above, $D[H(P)]$ can also be
calculated from the coefficients $C_{i}$ of the characteristic
polynomial without recourse to the roots. In the general case this
can be done since $D[H(P)]$ can be written in terms of the
determinant of the $(2n-1)\times(2n-1)$ Sylvester matrix of the
characteristic polynomial \cite{Basubook}:
\begin{equation}
\label{eq:Sylvester}  \small D[H(P)] =
\frac{(-1)^{\frac{n\left(n-1\right)}{2}}}{C_{n}}
\begin{vmatrix}
            C_{n}     & C_{n-1} & ...   & C_{0} &        &        &      \\
                      &  C_{n}  &       & ...   & C_{0}  &        &      \\
                      &         &       &       & ...    &        &      \\
                      &         &       &       &        & ...    & C_{0}\\
            nC_{n}    & ...     &       &       &        &        &      \\
                      & nC_{n}  & ...   & C_{1} &        &        &      \\
                      &         &       & ...   &        &        &      \\
                      &         &       &       &  .. .  & 2C_{2} & C_{1} \\
          \end{vmatrix}.
\end{equation}
Eq.(\ref{eq:Sylvester}) has two important implications. First, even
if the roots cannot be calculated analytically, which is true in
general for $n\geq 5$, $D[H(P)]$ always can be. This means that the
analytical mapping - supplied by the discriminant - between the
level-crossing and polynomial root-finding problems is preserved for
all $n$. Secondly, one can detect level crossings without solving
for the spectrum. This implies, as will be shown, the existence of
powerful algebraic alternatives to numerically intensive spectral
searches for crossings.

As in the case of $\Delta$ the real roots of Eqn.(\ref{eq:disc})
correspond to crossings in the spectrum of $H(P)$; as well, the real
parts of the complex roots of Eqn.(\ref{eq:disc}) indicate avoided
crossings \cite{Heiss1991}. Further, in our simple example, $\Delta$
turned out to be quadratic in a single parameter $T$; in general
$D[H(P)]$ will be a high order polynomial in $N$ variables.
Typically, in order to locate the roots of this polynomial we will
examine its coefficients. For example a simple method we will use is
Descartes' Rule of Signs from elementary algebra which equates the
number of real roots of a polynomial to the number of sign changes
in its coefficients, modulo 2. A more sophisticated concept we will
use is that of a Sturm-Habicht sequence \cite{Basubook} of a
polynomial, say in the variable $Y$. The difference in the number of
sign variations in this sequence at $Y = l$ and $Y = r$ equals the
number of real zeros of the polynomial in the interval $[l,r]$
exactly.

Let us now apply the methods outlined above to an atom in a uniform
magnetic field. Such a system is relevant to the search for a
Feshbach resonance, which typically begins with a spin-polarized
sample of atoms and requires a broad sweep of the magnetic field.
Experimentally, if a crossing or a fine anti-crossing is encountered
in the atomic spectrum while the magnetic field is tuned, some of
the atoms can be transferred to a different spin state, introducing
an impurity \cite{unanyan2001}. For theoretical bookkeeping, it is
useful to know if the low- and high-field states correlate
adiabatically \cite{bambini2002}. Such a correlation breaks down at
a crossing. For these reasons, it is essential to be aware of
crossing phenomena in the atomic spectrum. The magnetic fields
typically used in Feshbach resonance searches are typically less
than a KiloGauss. Hence the parameter regime of our interest will be
$B_{Fesh}\in$[0, 1 kG].

To work with a concrete example we consider an $^{23}$Na atom in its
$3S_{1/2}$ ground state split by the hyperfine interaction into two
levels labeled by the angular momentum $F=I+S$ where $I = 3/2$ and
$S=1/2$ are the nuclear and electronic spins respectively. In the
presence of a uniform magnetic field $B$ along the z-axis the atomic
spectrum is described by the Breit-Rabi Hamiltonian
\cite{breitrabi1931}
\begin{equation}
\label{eq:BreitRabiHamiltonian} H_{BR}=A  \textbf{I}  \cdot
\textbf{S} + B (a S_{z}+ b I_{z})
\end{equation}
where $A$ measures the strength of the hyperfine coupling and $a$
and $b$ may be easily related to the electronic and nuclear
gyromagnetic ratios and the Bohr magneton. Throughout this Letter we
will refer to the atomic eigenstates in the $|F,M_{F}\rangle$ basis
where $M_{F}$ is the component of $F$ along the z-axis. In order to
retain the polynomial form of the matrix elements, however, all
representations will be made in the uncoupled basis
$|M_{S},M_{I}\rangle$, where $M_{S,I}$ are the components of the
respective spins along the z-axis. Such a representation in the
ordered basis $|\frac{-1}{2},\frac{-3}{2}\rangle,
|\frac{1}{2},\frac{-3}{2}\rangle, |\frac{-1}{2},\frac{-1}{2}\rangle,
|\frac{1}{2},\frac{-1}{2}\rangle,|\frac{-1}{2},\frac{1}{2}\rangle$,
$|\frac{1}{2},\frac{1}{2}\rangle, |\frac{-1}{2},\frac{3}{2}\rangle$
and $|\frac{1}{2},\frac{3}{2}\rangle$ yields for
Eq.(\ref{eq:BreitRabiHamiltonian}):
\begin{widetext}
\begin{equation}
\label{eq:HBRmatrix} H_{BR} = \tiny \frac{1}{4}
\begin{pmatrix}
  3A-2(a+3b)B        &             &               &                       &              &             &             &   \\
                & -3A+2(a-3b)B     & 2\sqrt{3}A    &                       &              &             &             &   \\
                & 2\sqrt{3}A       &  A-2(a+b)B    &                       &              &             &             &   \\
                &                  &               & -A+2(a-b)B            & 4A           &             &             &   \\
                &                  &               & 4A                    & -A-2(a-b)B   &             &             &   \\
                &                  &               &                       &              & A+2(a+b)B   &2\sqrt{3}A   &   \\
                &                  &               &                       &              & 2\sqrt{3}A  &-3A-2(a-3b)B &   \\
                &                  &               &                       &              &             &             & 3A+2(a+3b)B \\
\end{pmatrix}
\end{equation}
\end{widetext}
To begin with a simplification we consider the fact that $\mid b
\mid\ll\mid a \mid$ in Eq.(\ref{eq:BreitRabiHamiltonian})
\cite{Metcalfbook}. Setting $b$ = 0 in Eq.(\ref{eq:HBRmatrix}) we
calculate the discriminant of $H_{BR}^{b=0}$ using built-in
functions in \emph{Mathematica}:
\begin{eqnarray}
\begin{array}{l}
\label{eq:BreitRabinob} D[H_{BR}^{b=0}]= \frac{81
A^{24}a^{26}B^{26}(64A^{6}+32A^{4}a^{2}B^{2}+8
A^{2}a^{4}B^{4}+a^{6}B^{6})}{65536}.\\
\end{array}
\end{eqnarray}
Considered as a polynomial in $B$, $D[H_{BR}^{b=0}]$ is even and
exhibits no sign changes in its coefficients; by Descartes' rule it
has no real roots for $B\neq0$ \cite{Basubook}. Hence there are no
level crossings in the spectrum. The real parts of the complex roots
of $D[H_{BR}^{b=0}]$ imply well documented, widely avoided crossings
at $B=0$ and $A/a$ (= 316 G for $^{23}$Na) \cite{Metcalfbook}.
Physically, the latter corresponds to the crossover into the
Paschen-Back regime.

We now calculate the discriminant $D[H_{BR}]$ for $b \neq 0$, the
expression for which will be presented elsewhere on account of its
length \cite{Crossingpra}. From the roots of $D[H_{BR}]$ we infer
the following changes introduced by the presence of $b$. The avoided
crossings are now located at $B=0$ and $A/(a-b)$ and have moved
closer since $b/a \sim -10^{-3}$ for $^{23}$Na. More importantly
there now appear crossings in the spectrum for $B\neq0$. Evidently
the coupling, $b$, of the nuclear spin to the magnetic field is the
mechanism responsible for crossings in the Breit-Rabi spectrum. For
$B>0$ there are 6 crossings clustered around 400 kG. In the regime
$B_{Fesh}\in$[0, 1 kG] therefore the discriminant alerts us to the
absence of any crossings as well as to the presence of two avoided
crossings.

To the best of our knowledge invariants of the Breit-Rabi
Hamiltonian such as Eq.(\ref{eq:BreitRabinob}), which can be
calculated once the spins $I$ and $S$ are specified, have not been
pointed out previously. They are complete catalogs of the parametric
symmetries of $H_{BR}$. More generally, although we have picked an
example from atomic physics, it should be clear that for any system
represented by a parameter-dependent matrix, the discriminant is a
powerful tool for investigating the location of crossings, their
behavior as a function of the parameters as well as the physical
mechanisms responsible for their occurrence.

We now show how this method yields quite remarkable results in the
more involved case of a diatomic molecule in a magnetic field. We
consider a molecule made up of the previously considered $^{23}$Na
atom and an $^{85}$Rb atom. The molecular Hamiltonian is
\begin{equation}
\label{eq:Hmol} H_{Mol} = K.E.+U(B)+V(R).
\end{equation}
Here $K.E.$ denotes the kinetic energy of the nuclei, $U(B)$ the
internal energies of the atoms in the magnetic field, and $V(R)$ the
molecular potential energy at the internuclear distance $R$.
Specifically, $U(B)$ can be obtained from
Eq.(\ref{eq:BreitRabiHamiltonian}), and
\begin{equation}
\label{eq:molpot1} V(R)=V_{S}P_{S}+V_{T}P_{T},
\end{equation}
where $V_{S,(T)}$ are the molecular (singlet, triplet) electrostatic
potentials, and $P_{S,T}$ are projection operators on to the
corresponding subspaces \cite{bambini2002}.

To work with a concrete example we prepare the Na-Rb atom pair in
the state $|2,2\rangle_{Na} |2,2\rangle_{Rb}$, the ``open'' channel
for their collision. Since the component of the total angular
momentum ($M_{Ftotal}$ = 4) along the magnetic field is preserved by
the interaction $V(R)$, we need to consider in addition only the
states which have the same $M_{Ftotal}$: $|2,2\rangle|3,2\rangle$,
$|2, 1\rangle|3,3\rangle $, and $|1,1\rangle|3,3\rangle$. Here we
have retained the ordering of the atoms and dropped the subscripts.
The four states mentioned above make up the electronic Hilbert space
of the molecule.

In the standard Born-Oppenheimer approximation (BOA) of molecular
physics the $K.E.$ in Eq.(\ref{eq:Hmol}) is at first neglected and a
set of adiabatic electronic potential curves obtained by
diagonalizing the (in our case $4\times4$) interaction matrix
\begin{equation}
H_{BO} = U(B)+V(R),
\end{equation}
for a given $B$ and for a large enough range of $R$. In order to
find the bound rovibrational levels one must then solve for the
nuclear motion on these curves. The calculation has to be repeated
for different values of $B$ until a bound state equals the open
channel in energy : this yields the simplest estimate of a Feshbach
resonance location \cite{bambini2002}.

It is imperative to know if there are crossings among the electronic
potentials generated by $H_{BO}$, because the BOA breaks down
precisely at such points, and invalidates the Feshbach resonance
estimate. For this reason we investigate the discriminant
$D[H_{BO}]$.  But first we make a crucial transformation. We rewrite
Eq.(\ref{eq:molpot1}) as
\begin{equation}
\label{eq:molpot2}
V(R)= X(P_{S}-P_{T}) + (V_{S}+V_{T})/2,
\end{equation}
where $X(R)=(V_{S}-V_{T})/2$, signifies the quantum mechanical
exchange energy responsible for chemical bonding and plays a
critical role in the curve crossing problem by allowing us to avoid
the following difficulty. The $V_{S,T}$ depend very sensitively on
$R$. For $^{23}$Na$^{85}$Rb for instance, the latest data requires
32(47) parameters to fit $V_{S(T)}$ respectively \cite{pashov2005}.
These parameters are usually varied within their uncertainties, for
instance to yield bounds on the Feshbach resonance predictions. It
would require a numerical effort in (32+47+1=)80-dimensional
parameter space, always limited by resolution, to inspect the
spectrum for crossings for each $V_{S},V_{T}$ and $B$. We
demonstrate below how the introduction of $X$ renders the
determination of curve crossings for $H_{BO}$ completely insensitive
to the complicated functional form of the Born-Oppenheimer
potentials.

Since $D[H_{BO}]$ does not change if we subtract the same quantity
from all the diagonal elements of $H_{BO}$, we drop the second term
in Eq.(\ref{eq:molpot2}), i.e. we calculate
$D[H_{BO}-(V_{S}+V_{T})/2] (=D[H_{BO}])$. We then substitute numbers
for all the \emph{atomic} constants in $D[H_{BO}]$ which gives us
$D[H_{BO}]=\sum_{n=0}^{6} p_{n}(B)X^{n}$ i.e., a polynomial of
degree 6 in $X$, whose coefficients depend on $B$. We now phrase our
enquiry about curve crossings in the following way : is there
\emph{any} real value of the exchange energy
($X\in[-\infty,+\infty]$) for which molecular curve crossings can be
produced ($D[H_{BO}]=0$) by the fields available in the laboratory
($B_{Fesh}\in$[0,1 kG]) ? The answer can be found by examining the
number of sign variations in the Sturm-Habicht sequence of
$D[H_{BO}]$ at $X =\pm\infty$ \cite{Basubook}. Only two real roots
are found in the interval $B_{Fesh}\in$[0,1 kG], at $B$ = 0G and
502.2G, both for $X=0$. Other than these we locate no crossings for
any $X$ and the BOA is valid everywhere else in the parameter regime
of interest.

We think it is quite remarkable, and points to the power of the
algebraic approach, that in order to extract information about
crossings in the spectrum of a diatomic molecule in a magnetic field
we did not avail of any knowledge about the Born-Oppenheimer
potentials ($V_{S,T}$) except that their difference is a real
number! The curve crossings are entirely determined by atomic
constants. Of course whether and at what $R$ the crossings actually
occur will depend on the specific form of the $V_{S,T}$. The
algebraic method allows us to locate the complete set of magnetic
fields that could cause the BOA to break down. If this set is not
too restrictive, or is not in the neighborhood of a Feshbach
resonance, one need not refer to the $V_{S,T}$ at all.

Having demonstrated the application of curve-crossing methods to
atomic and molecular spectra we now discuss possible extensions of
our work. Our methods can readily be applied to other atoms and
molecules. More generally we note that
Eq.(\ref{eq:BreitRabiHamiltonian}) is just a single example from an
entire class of ``spin-Hamiltonians''
\begin{equation}
H_{spin} =\sum_{i,j} C_{ij}J_{i}J_{j}+\sum_{i} B_{i}J_{i}.
\end{equation}
Here the $J_{i}$ are angular momentum operators and the indices
$i,j$ run over the spatial coordinates $x,y,z$. With the appropriate
choices for the `coupling' ($C_{ij}$) and `field' ($B_{i}$)
parameters, effective Hamiltonians for many physical systems can be
set up and analyzed for level crossings. Specific examples of such
systems are polar molecules in an electric field \cite{Hain1999}, a
spin-1 BEC with magnetic dipole interaction \cite{HanSpinor},
doubly-even nuclei described by the triaxial rotor model
\cite{wood2004}, massive spinors in an anisotropic universe
\cite{dowker1974}, and so on. Of course our methods are not limited
to Hamiltonian matrices. They can be applied to matrices that
describe the stability of gyroscopes \cite{Seyranianbook}, the
polarization of light \cite{freund2004}, or the payoff in a game
\cite{Petrosjanbook}, for instance. In sum, a large number of
mathematical models contain parameters, and many of them involve
matrices. We have shown how to extract curve crossing information
from cases with effectively one and two tunable parameters.
Application to other systems as well as to a larger number of
parameters are among the natural extensions of our work.

In conclusion we have demonstrated a powerful and rigorous algebraic
method for locating curve crossings in the spectra of physical
systems. Along the way we have pointed out a new class of invariants
of the Breit-Rabi equation of magnetic resonance and placed the
Born-Oppenheimer approximation for calculating Feshbach resonances
on a rigorous basis, using very little information about molecular
potentials.

\begin{acknowledgments}
M. B. would like to thank Dr. S. R. Muniz for encouraging him to
pursue this research, and Dr. L. Baksmaty and Prof. S. Basu for
technical help. This research was funded by DoE and ARO.
\end{acknowledgments}

\end{document}